\documentclass{iopart}
\usepackage{graphicx}
\usepackage{bm}

\begin{document}
\title{Development of simulation package \lq ELSES'
for extra-large-scale electronic-structure calculation}
\author{T. Hoshi$^{1,2}$, T. Fujiwara$^{3,2}$}
\address{(1) Department of Applied Mathematics and Physics, 
Tottori University, Tottori 680-8550, Japan}
\address{(2) Core Research for Evolutional Science and Technology, 
Japan Science and Technology Agency (CREST-JST), Japan}
\address{(3) Center for Research and Development of Higher Education, 
The University of Tokyo, 
Bunkyo-ku, Tokyo, 113-8656, Japan}
\begin{abstract}
An early-stage version of simulation package \lq ELSES' 
(Extra-Large-Scale Electronic-Structure calculation)
is developed 
for electronic structure and dynamics of large systems, 
particularly, nm-scale or 10nm-scale systems 
(www.elses.jp). 
Input and output files are written in 
the Extensible Markup Language (XML) style for general users.
Related pre-/post-simulation tools are also available. 
Practical work flow and example are described. 
A test calculation of GaAs bulk system is shown to
demonstrate that the present code can handle systems with more than one atom species.
Several future aspects are also discussed. 
\end{abstract}

\date{\today}
\maketitle
\section{Introduction}

\begin{figure}[htbp] 
\begin{center}
\resizebox{0.95\textwidth}{!}{
  \includegraphics{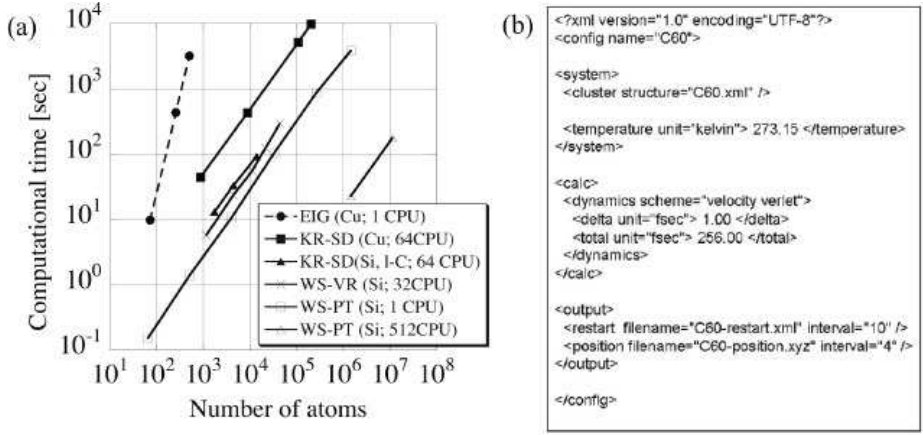}
}
\end{center}
\caption{\label{FIG-BENCH} 
(a) The computational time
as a function of the number of atoms ($N$). 
\cite{HOSHI2003, HOSHI2005, HOSHI2006}
The time was measured 
for metallic (fcc Cu and liquid C) and insulating (bulk Si) 
systems with up to 11,315,021 atoms, 
by the conventional eigenstate calculation (EIG) 
and by our methods for large systems; 
Krylov-subspace method with subspace diagonalization procedure (KR-SD), 
and Wannier-state method with variational and perturbative procedures
(WS-VR, WS-PT).  
See the original papers~\cite{HOSHI2003, HOSHI2005, HOSHI2006}
for the details of parallel computation.
(b) An example of XML formatted input file. 
 In this example of input file, calculation conditions are set as the initial atomic structure file: C60.xml,
integration algorithm for molecular dynamic calculation: velocity Verlet, time step: 1.0fsec, total simulated time:
250fsec. Results of atomic structures are written in a file named  \lq C60-position.xyz ' every 4 steps, and restart file
\lq C60-restart.xml' is updated every 10 steps.
}
\end{figure}

Process (molecular-dynamics, MD) calculation with electronic structure is essential
as analysis or prediction tools of nanomaterials, 
particularly, materials in  nanometer or ten-nanometer scales.
Since structure and function of materials in these scales are determined
by the competition among different regions, typically surface and bulk (inner) regions,
the theory should reproduce such a nanoscale competition by describing 
correct electronic structures at different regions.
For years, we have developed a set of theories and program code for
such nanoscience researches. 
\cite{HOSHI2000, HOSHI2003, geshi2004, TAKAYAMA2004, 
HOSHI2005, TAKAYAMA2006, HOSHI2006, IGUCHI2007, FUJIWARA2008} 
One crucial point is
that large-scale quantum-mechanical calculation can be realized, in principle, by calculating the one-body density
matrix, instead of one-electron eigen states, since the computational cost can be drastically reduced. \cite{KOHN96}
An overview of these theories can be found in the introduction part of Ref.~\cite{HOSHI2006}.
Practical methods were constructed as solver methods of the one-body density matrix or the Green's function for a given
Hamiltonian matrix.  We note that some of the theories are purely mathematical ones,
iterative linear-algebraic algorithms for large matrices and, therefore, should be useful in other fields of physics.
Actually, one method, called 'shifted conjugate-orthogonal conjugate-gradient method', 
\cite{TAKAYAMA2006} was applied to an
extended Hubbard model for La$_{2-x}$Sr$_x$Ni$_2$O$_4$.
\cite{YAMAMOTO2007}
Another crucial point is to construct algorithms for efficient
parallel computations. Since multi-core CPU architectures are now built in standard workstation or personal
computer, parallel computations are essential for actually all the computational systems. The calculations are
realized with Slater-Koster-form (tight-binding) Hamiltonians and test calculation was carried out 
with 10$^2$-10$^7$ atoms with or without parallelism.  
See Fig.\ref{FIG-BENCH} (a) for a bench mark,
in which the parallelism was realized by the OpenMP directive (www.openmp.org).
As a benchmark with a recent multi-core CPU architecture, we
have tested our code with a standard workstation with four dual-core CPU's (Opteron 2GHz), for liquid carbon
with 1728 atoms. We adopted a typical Hamiltonian of carbon system. \cite{XU} 
As a result, a computational time is
six seconds per time step in the process (MD) calculations and a parallel efficiency is more than 90\%. 
Electronic property, such as density of states, is also calculated. \cite{HOSHI2006}

Now the code has named Extra Large Scale Electronic Structure calculation (ELSES) code (www.elses.jp).
It is reorganized as a simulation package with  input/output files 
in the Extensible Markup Language (XML) style (http://www.w3.org/XML/),
such as shown in Fig.\ref{FIG-BENCH} (b),  
for a wider range of users and applications. 
This article describes  structure, example and future aspect of the simulation code .

\begin{figure}[ht]
 \begin{center}
  \includegraphics*[width = 0.6 \textwidth]{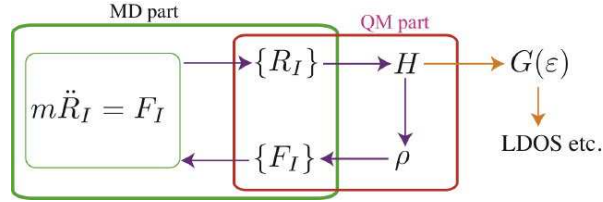}
 \end{center}
 \caption{Illustrated structure of the code   \label{FIG-OVERVIEW2}
}
\end{figure}


\section{Structure of code}


Structure of the code is illustrated schematically 
in Fig.~\ref{FIG-OVERVIEW2}.
The outer loop of the code 
is the loop of time evolution in MD simulation (MD part)
according to the Newton equation;
\begin{eqnarray}
M_I \ddot{\bm{R}}_I =  \bm{F}_I 
\end{eqnarray}
where $M_I,\bm{R}_I$ are 
mass and position of $I$-th atom (ion), respectively
and $ \bm{F}_I $ is the force on the atom.
The time evolution is performed numerically
by a finite-difference method,
as in classical MD simulations.
At each time step, 
the quantum-mechanics (QM) part is called from the MD part,
in which the force is calculated from the position of atoms;
\begin{eqnarray}
 \{ \bm{R}_I \} \Rightarrow H \Rightarrow \rho \Rightarrow \{ \bm{F}_I \}.
\end{eqnarray}
Here $H$ is Hamiltonian matrix and $\rho$ is the one-body density matrix,
defined formally as
\begin{eqnarray}
\rho(\bm{r}, \bm{r}') \equiv \sum_i f_i \phi_i^*(\bm{r}) \phi_i(\bm{r}'),
\end{eqnarray}
from eigen states $\{ \phi_i(\bm{r}) \}$
and their occupation number $\{ f_i \}$.
The diagonal elements ($\bm{r} = \bm{r}'$)
gives electron density ($n(\bm{r})\equiv \rho(\bm{r}, \bm{r})$)
and the off-diagonal ones ($\bm{r} \ne \bm{r}'$) 
are responsible for quantum-mechanical effect.
In an order-$N$ calculation, such as Krylov-subspace method
\cite{TAKAYAMA2004, HOSHI2006},
the density matrix is calculated without eigen states $\{\phi_i(\bm{r})\}$. 
In the present article, 
the calculation method of the density matrix from Hamiltonian
($H \Rightarrow \rho$)
is called \lq solver'.  
In the practical code,
the matrices are given within atomic orbital representation
\begin{eqnarray}
H(\alpha,\beta,I,J) =\langle \alpha, I | \hat{H} | J \beta \rangle \\
\rho(\alpha,\beta,I,J) =\langle \alpha, I | \hat{ \rho}  | J \beta \rangle
\end{eqnarray}
where the suffices $I,J$ denote atoms and
the suffices $\alpha,\beta$ denote orbitals.
The local density of states (LDOS) 
is calculated through the Green function $G(\varepsilon)$,
with a Krylov-subspace solver  \cite{TAKAYAMA2006};
\begin{eqnarray}
H \Rightarrow G \Rightarrow {\rm LDOS},
\end{eqnarray}
which is carried out in a post-simulation tool.


\begin{figure}[ht]
 \begin{center}
  \includegraphics*[width = \textwidth]{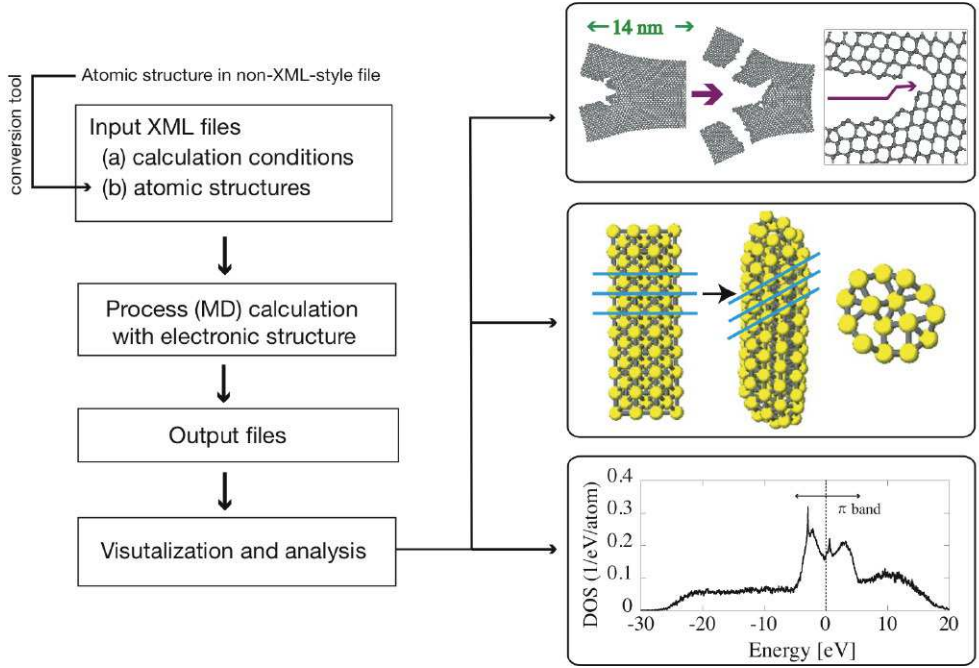}
 \end{center}
 \caption{  \label{FIG-APPLI}
Schematic figure of the work flow of research with examples of results (right panels). 
As examples, silicon cleavage process, \cite{HOSHI2005}
formation process \cite{IGUCHI2007} of helical multishell gold nanowire \cite{TAKAYANAGI2000},
local density of state of liquid carbon 
\cite{HOSHI2006} appear,
at the upper, middle and lower panels of the right side, respectively.}
\end{figure}


Figure \ref{FIG-APPLI} indicates the work flow in research;
Two input files in the XML style are needed; 
(a) file calculation conditions (See Fig.~\ref{FIG-BENCH}(b)) 
and (b) file for initial atomic structure. 
Several conversion tools, as pre-simulation tools,  
are available so as to create 
the XML-style file of
atomic structure from non-XML style files, such as files in the conventional XYZ style.

 The XML style is used commonly among text-based electronic files 
for sharing data through the internet.
It is flexible and extensive, 
because it allows us to define their own items.
In an XML-style file,  
each item is designated by start tag ($< \cdot \cdot \cdot >$) and end tag ($<$/$\cdot \cdot \cdot>$). 
As an example of our XML-style file,   Fig.~\ref{FIG-BENCH}(b) contains a line of
\lq $<$temperature unit="kelvin"$>$273.15 $<$/temperature$>$',
which means that the temperature of the system is set to be $T$=273.15 Kelvin. 
Other units can be used by rewriting the \lq unit' part, such as \lq unit="eV" ',  for eV. 

We note that the
implementation of a XML-style input file is important in practical simulations of
nanostructure materials, since various conditions are required. 
For example, the fracture
simulation of silicon nanocrystal \cite{HOSHI2005} was realized by imposing an external load 
on the atoms in a limited region
near the sample boundary. The extendibility of XML-style file can satisfy these detailed conditions, by adding
newly-defined tags for its own purpose. 

After the MD simulation, 
we analyze the atomic processes and electronic structure
in detail as well as to visualize atomic structures. 
The LDOS calculation tool is available as a post-simulation tool,
as discussed above. 
The tools should be developed further, particularly,
for analyzing electronic structure,
such as crystal orbital Hamiltonian populations \cite{COCG, TAKAYAMA2006}, 
a quantitative visualization method of  a chemical bond 
from the energetics with off-site elements of the Green's function $G(\varepsilon)$.

\section{Test calculation of compound; example of bulk GaAs}

Bulk GaAs was calculated so as to demonstrate that the present code can handle systems with more than one
atom specie. We adopt a Slater-Koster-form Hamiltonian of GaAs with s, p and s* atomic orbitals. 
\cite{MOLTENI}
The atomic energy level of the s* orbital is located within the conduction band and its physical origin is a spherical
average of the five d orbitals. The formulation of s, p, and s* orbitals was introduced \cite{VOGL} among various
semiconductors, for reproducing the valence band and the bottom of the conduction band and was used in papers,
such as Refs. ~\cite{MOLTENI, SHKREBTII, MOLTENI2, SEONG, CAPAZ, SANTOPRETE, VOLPE},
for liquid, amorphous, defect, surface and quantum dot.
Figure 4 shows calculation results of bulk GaAs, in which the cubic periodic cell with 64 atoms is used. Here the
Krylov-subspace method with subspace diagonalization \cite{TAKAYAMA2004,HOSHI2006} is adopted 
for solver routine of the density matrix.
The dimension of the Krylov subspace (Krylov dimension) should be set as a controlling parameter that
determines accuracy and computational cost. The computational time is proportional to the Krylov dimension and
the calculation will be converged to the exact one, when the Krylov dimension increases. 
See Ref.~\cite{HOSHI2006} for detail.
Figure 4 plots the optimized lattice constant and the energy as the function of the Krylov dimension. Figure 4
indicates that the calculation is well converged with the Krylov dimension of 30; the deviations in the lattice
constant and the energy are less than 0.01 \% 
and less than 1meV per atom, respectively. We note that an excellent
convergence at the Krylov dimension of 30 was found in the other systems. 
\cite{TAKAYAMA2004,HOSHI2006}

\begin{figure}[htbp] 
\begin{center}
\resizebox{0.48\textwidth}{!}{
  \includegraphics{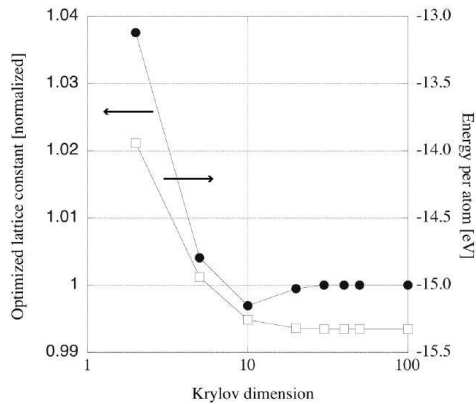}
}
\end{center}
\caption{\label{FIG-GAAS} 
The optimized lattice constant 
and the energy for bulk GaAs 
as the function of the Krylov dimension. 
}
\end{figure}

\section{Summary}

Large-scale electronic structure calculation code is being developed as a simulation package with the name of
ELSES (www.elses.jp). For a better user interface of our simulation code, we have created the input/output
interfaces of XML-style files. The pre- and post-processing tools have been also prepared for modeling and
detailed analysis of the atomic structures. We have also confirmed that the present code can handle system with
more than one atom specie by calculating bulk GaAs. Although the present stage of the simulation package is still
in an early one, we believe that the code will provide fruitful simulations for researchers in nano-material science.
As a future aspect, 
the method is being extended by implementation of  
general Slater-Koster-form Hamiltonians 
for wider range of materials, in which 
an explicit charge selfconsistent  treatment \cite{ELSTNER} is included. 
Non-equilibrium current and other electronic properties are also crucial for nanoscience and 
should be investigated with the present methodologies.

\section*{Acknowledgments}

Numerical calculation was partly carried out 
using the supercomputer facilities of
the Institute for Solid State Physics, University of Tokyo and
the Research Center for Computational Science, Okazaki. 

\section*{References}



\begin{thebibliography}{}


\bibitem{HOSHI2000}
T. Hoshi and T. Fujiwara, 
J. Phys. Soc. Jpn. {\bf 69}, 3773 (2000); \\
Preprint: http://arxiv.org/abs/cond-mat/9910424

\bibitem{HOSHI2003}
T. Hoshi and T. Fujiwara, 
J. Phys. Soc. Jpn. {\bf 72}, 2429 (2003); \\
Preprint: http://arxiv.org/abs/cond-mat/0210366

\bibitem{geshi2004}
M. Geshi, T. Hoshi and T. Fujiwara, J. Phys. Soc. Jpn., {\bf 72}, 2880 (2003); \\
Preprint: http://arxiv.org/abs/cond-mat/0306461

\bibitem{TAKAYAMA2004}
R. Takayama, T. Hoshi and T. Fujiwara, 
J. Phys. Soc. Jpn. {\bf 73}, 1519 (2004); \\
Preprint: http://arxiv.org/abs/cond-mat/0401498

\bibitem{HOSHI2005}
T. Hoshi, Y. Iguchi and T. Fujiwara, 
Phys. Rev. B{\bf 72}, 075323 (2005); \\
Preprint: http://arxiv.org/abs/cond-mat/0611738

\bibitem{TAKAYAMA2006}
R. Takayama, T. Hoshi, T. Sogabe,  S-L. Zhang and T. Fujiwara, 
Phys. Rev. B{\bf 73}, 165108 (2006); \\
Preprint: http://arxiv.org/abs/cond-mat/0503394

\bibitem{HOSHI2006}
T. Hoshi,  and T. Fujiwara, 
J. Phys: Condens. Matter. {\bf 18}, 10787 (2006); \\
Preprint: http://arxiv.org/abs/cond-mat/0610563

\bibitem{IGUCHI2007}
Y. Iguchi, T. Hoshi and T. Fujiwara, 
Phys. Rev. Lett. {\bf 99}, 125507 (2007); \\
Preprint: http://arxiv.org/abs/cond-mat/0611738

\bibitem{FUJIWARA2008}
T. Fujiwara, T. Hoshi and S. Yamamoto,
J. Phys: Condens. Matter. {\bf 20}, 294202 (2008);  \\
Preprint: http://arxiv.org/abs/0802.0748

\bibitem{KOHN96}
W. Kohn, Phys. Rev. Lett. {\bf 76}, 3168 (1996).

\bibitem{YAMAMOTO2007}
S. Yamamoto, T. Fujiwara and Y. Hatsugai, Phys. Rev. B76, 165114 (2007); \\
Preprint: http://arxiv.org/abs/0704.3323

\bibitem{XU}
C. H. Xu, C. Z. Wang, C. T.  Chan and K. M. Ho,  
J. Phys. Condens. Matter {\bf  4},  6047 (1992).



\bibitem{TAKAYANAGI2000}
Y. Kondo and K. Takayanagi, Science {\bf 289}, 606 (2000).


\bibitem{COCG}
 R. Dronskowski and P. E. Bl\"ochl, J. Phys. Chem. {\bf 97}, 8617 (1993).

\bibitem{MOLTENI}
C. Molteni, L. Colombo and L. Miglio, J. Phys. Condens. Matter 6, 5243 (1994); ibid, 5257 (1994).

\bibitem{VOGL}
P. Vogl, H. P. Hjalmarson and J. D. Dow, J. Phys. Chem. Solids 44, 365 (1983).

\bibitem{SHKREBTII}
A. I. Shkrebtii and R. Del sole, Phys. Rev. Lett. 70, 2645 (1993).

\bibitem{MOLTENI2}
C. Molteni, L. Colombo and L. Migilio, Phys. Rev. B50, 4371 (1994).

\bibitem{SEONG}
H. Seong and L. J. Lewis, Phys. Rev. B52, 5675 (1995).

\bibitem{CAPAZ}
R. B. Capaz, K. Cho and J. D. Joannopoulos, Phys. Rev. Lett. 75, 1811 (1995).

\bibitem{SANTOPRETE}
R. Santoprete, B. Koiller, R. B. Capaz, P. Kratzer, Q. K. K. Liu and M. Scheffler, Phys. Rev. B68, 235311
(2003).

\bibitem{VOLPE}
M. Volpe, G. Zollo and L. Colombo, Phys. Rev. B71, 075207 (2005).

\bibitem{ELSTNER}
M. Elstner, D. Porezag, G. Jungnickel, J. Elsner, M. Haugk, Th. Frauenheim, S.
Suhai and G. Seifert, Phys. Rev. B{\bf 58}, 7260 (1998).

\end{thebibliography}
\end{document}